\documentstyle[twocolumn,epsbox]{jpsj}

\def\runtitle{Singlet Ground State and Magnetization Plateaus in Ba$_3$Mn$_2$O$_8$}
\def\runauthor{Masahiro {\sc Uchida} {\it et al.}}
\newcommand{\mol}[1]{$\rm{#1}$}

\newcommand{\green}{\mol{Ba_3Mn_2O_8}}

\title
{Singlet Ground State and Magnetization Plateaus in Ba$_3$Mn$_2$O$_8$
}

\author
{ 
Masahiro {\sc Uchida}\footnote{E-mail address: uchida@lee.phys.titech.ac.jp.}, Hidekazu {\sc Tanaka}, Mikhail {\sc Bartashevich}$^1$\\ 
and Tsuneaki {\sc Goto}$^1$
}

\inst
{
Department of Physics, Tokyo Institute of Technology, Oh-okayama, Meguro-ku, Tokyo 152-8551\\
$^1$Institute for Solid State Physics, The University of Tokyo, Kashiwanoha, Kashiwa, Chiba 277-8581
}

\recdate
{
\hspace{7cm}
}

\abst
{
Magnetic susceptibility and the magnetization process have been measured in \green\  polycrystal. In this compound the magnetic manganese ion exists as Mn$^{5+}$ in a tetrahedral environment, and thus the magnetic interaction can be described by an $S=1$ Heisenberg model. The ground state was found to be a spin singlet with an excitation gap $\Delta/k_{\rm B}=11.2$ K. Magnetization plateaus were observed at zero and at half of the saturation magnetization. These results indicate that the present system can be represented by a coupled antiferromagnetic dimer model. 
}

\kword
{
Ba$_3$Mn$_2$O$_8$, Mn$^{5+}$, $S=1$ quantum spin system, singlet ground state, excitation gap, magnetic susceptibility, high-field magnetization process, magnetization plateaus
}

\begin{document}
\sloppy
\maketitle

\section{Introduction}
Tribarium dimanganese octaoxide, \green , consists of \mol{Ba^{2+}} cations and \mol{MnO^{3-}_4}anions.\cite{Wyckoff} 
In this compound, the transition metal Mn exists as \mol{Mn^{5+}}.
The oxidation state is very rarely found in stable oxides, although it is known that manganese has various oxidation states, {\it e.g.}, Mn$^{3+}$ in LaMnO$_3$, Mn$^{4+}$ in CaMnO$_3$ and Mn$^{7+}$ in KMnO$_4$.

Recently, the crystal structure of \green\ was determined precisely by X-ray and neutron powder diffractions.\cite{Weller} The space group of this compound is trigonal \mol{R\bar{3}m}. The lattice constants are $a=5.71\ \rm{\AA}$ and $c=21.44\ \rm{\AA}$. \mol{Mn^{5+}} ions are located at the center of tetrahedra of \mol{O^{2-}} ions. All of the \mol{Mn^{5+}} sites are equivalent. The \mol{Mn^{5+}} ions form double-layered triangular lattices in the basal plane, which are stacked along the $c$-axis with a periodicity of three, as shown in Fig. 1. If the exchange interactions $J_1$ and $J_2$ are antiferromagnetic, they can produce spin frustration. Therefore, \green\ is an interesting system from both chemical and physical points of view.
\begin{figure}[htpb]
    \epsfigure{file=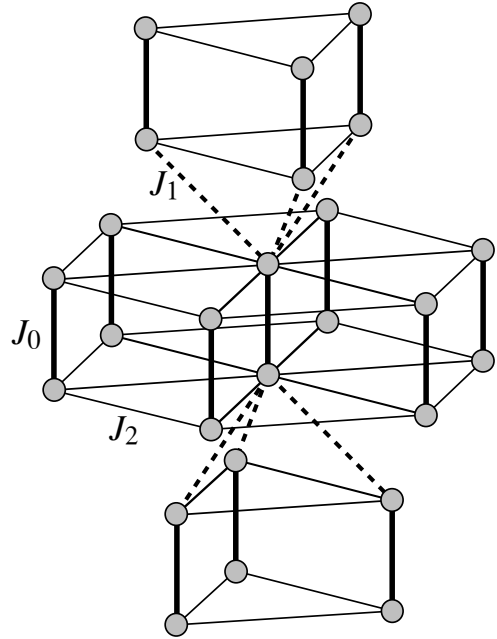,width=6.5cm}
    \caption{Arrangement of \mol{Mn^{5+}} ions which have spin 1 in \green. Thick, dashed and thin lines denote the first-, the second- and the third-nearest-neighbor exchange interactions $J_0$, $J_1$ and $J_2$, respectively.}
    \label{Network}
\end{figure}
   
The precise magnetic measurements of \green\ have not been reported so far. In this paper, we present the results of magnetic susceptibility and high-field magnetization measurements for purified \green\ polycrystal. As shown later, a gapped ground state is intrinsic to the present system.

\section{Experimental Procedures}
\green\ polycrystal was prepared according to the chemical reaction
$$\rm{
3BaCO_3 + 2MnO_2 + \frac{1}{2}O_2 \rightarrow Ba_3Mn_2O_8 + 3CO_2 .
}$$
In this reaction, \mol{Mn^{4+}} is oxidized to \mol{Mn^{5+}}. 
Reagent-grade \mol{BaCO_3} and \mol{MnO_2} were mixed in stoichiometric quantities, and calcined at 900 $^{\circ}$C for 30 hours in air. A greenish material obtained was examined by X-ray powder diffraction, and the findings were compared with reported data.\cite{Mizutani}
Although the diffraction pattern obtained has the features peculiar to \green, many unsystematic peaks due to impurities were observed.
Magnetic susceptibility measurements of this material revealed the existence of a small ferromagnetic impurity phase with a transition temperature $T_{\rm c}=43\rm{K}$, which seems to be \mol{Mn_3O_4}.\cite{Romanov}

To obtain high-purity \green, the material was sintered for more than 30 hours after being crushed and pressed into a pellet.
The same process was repeated four times, and the total sintering time was more than 120 hours.
The sample obtained finally was examined again by X-ray powder diffraction. No distinct peak due to impurities was observed. 
In the magnetic susceptibility data, no anomaly indicating ferromagnetic phase transition was observed at 43 K. All the experimental results presented in this paper were obtained using this sample.

The susceptibility of the polycrystal sample was measured between 1.8 K and 300 K at $H=0.1$ T using a SQUID magnetometer (Quantum Design MPMS XL). The magnetization measurement was performed using an induction method with a multilayer pulse magnet at the Ultra-High Magnetic Field Laboratory, Institute for Solid State Physics, The University of Tokyo. The powdered sample with the volume of $\sim$0.07 cm$^3$ was packed into a sample holder. Magnetization data were collected at 1.4 K in magnetic fields up to 47 T. 

ESR measurement at X ($\sim 9$ GHz) band frequencies was performed to obtain the $g$-factor. A single ESR absorption with the linewidth of $\sim$10 mT was observed. From the resonance field, the $g$-factor was determined to be $g=1.98$.

\section{Results and Discussions}
Figure 2 shows the magnetic susceptibility of \green\ as a function of temperature. Solid circles denote the raw data. When the temperature is decreased, the susceptibility exhibits a broad maximum at $T_{\rm max}=18$ K, and then decreases rapidly. The inset shows the low-temperature susceptibility of \green. No sharp anomaly indicative of the phase transition is seen. The susceptibility has a rounded minimum at 2.5 K and then increases again. The susceptibility behavior is not intrinsic, but due to a small amount of impurities or lattice defects, because the minimum vanishes with increasing applied magnetic field. Assuming that the susceptibility due to the impurity phase obeys the Curie-Weiss law, we subtracted it from the raw data. The net susceptibility was renormalized per mole and plotted using open circles in Fig. 2. From the susceptibility results, we can conclude that the magnetic ground state of \green\ is a spin singlet with an excitation gap (spin gap). 
\begin{figure}[htpb]
    \epsfigure{file=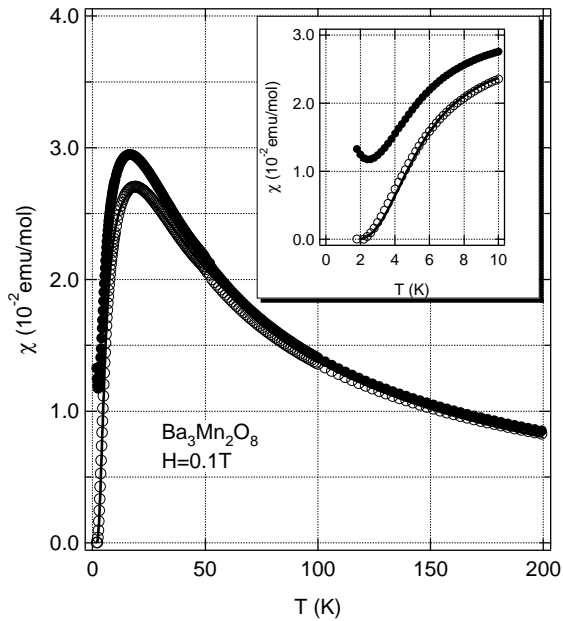,width=8cm}
    \caption{Temperature dependence of the magnetic susceptibility of \green\ measured at $H=0.1\rm{T}$. Solid circles denote the raw data. The susceptibility corrected for the Curie-Weiss term due to impurity phase is plotted using open circles. The inset shows the low-temperature susceptibility. The solid line is the fit obtained for eq. (2).}
    \label{susceptibility}
\end{figure}   

To understand the magnetic properties of \green , it is important to consider the crystal field acting on \mol{Mn^{5+}} and its orbital state. From Hund's rules, the electronic ground state of \mol{Mn^{5+}} ion with the 3$d^2$ configuration is $^3F$. In the presence of a cubic tetrahedral crystal field, this state splits into three states ($\Gamma_4$, $\Gamma_5$ and $\Gamma_2$). The ground state is the nondegenerate $\Gamma_2$ state. Consequently, the magnetic moment is approximately given by spin only. For this reason, the magnetic interactions in \green \ should be described by an $S=1$ Heisenberg model ${\cal H}=\sum_{\langle i,j\rangle }2J_{ij}{\mib S}_i\cdot {\mib S}_j$.
 
Next, we consider the origin of the gapped ground state. In Fig. 1 the first-, the second- and the third-nearest-neighbor exchange interactions, $J_0$, $J_1$ and $J_2$, are represented by thick, dashed and thin lines, respectively. Since $J_0$ is the interaction between the nearest neighbors, it appears dominant. A spin interacts with three and six other spins through $J_1$ and $J_2$ interactions, respectively. If $J_1$ or $J_2$ is dominant and a spin singlet state is formed on a bond coupled through either of them, the other two or five bonds cannot gain the exchange energies. Consequently, it appears impossible that the $J_1$ or $J_2$ interaction produces the spin gap. For these reasons, we can assume that the $J_0$ interaction is antiferromagnetic and dominant, and that the spin dimer coupled through $J_0$ is the origin of the spin gap.

The total magnetic susceptibility of isolated $S=1$ dimers, which are coupled through $J_0$, is expressed as 
$$
\chi_0 =  \frac{2N \beta g^2 \mu_{\rm B}^2 \left(1+5e^{-4\beta J_0}\right)}
		{3+e^{2\beta J_0}+5e^{-4\beta J_0}} ,
		\eqno(1)
$$
where $N$ is the number of dimers and $\beta=1/k_{\rm B}T$. The temperature dependence of the susceptibility observed is qualitatively described by eq. (1). However, there is some disagreement between the experimental data and eq. (1), and the $g$-factor $g=1.52$ obtained by the fitting is fairly small compared with $g=1.98$ obtained by the ESR measurements. Therefore, we should consider interdimer interactions $J_1$ and $J_2$. When these interactions are included in the calculation as effective fields, the susceptibility $\chi$ can be written as
$$
\chi =  \frac{\chi_0}{1+{\gamma}\chi_0} ,
		\eqno(2)
$$
with $\gamma=3(J_1 + 2J_2)/Ng^2 \mu_{\rm B}^2$.
Using the $g$-factor $g=1.98$ obtained by ESR measurements, we fit eq. (2) to the susceptibility data between 6 K and 300 K, and obtain $J_0/k_{\rm B}=9.2$ K and $(J_1+2J_2)/k_{\rm B}=6.6$ K. The fitting is almost perfect as shown by the solid line. 
We see that the intradimer interaction $J_0$ and the total of the interdimer interactions $3J_1+6J_2$ are of the same order of magnitude.
From the susceptibility results, we conclude that the present system is a coupled antiferromagnetic dimer system.

In order to investigate the phase transitions in a magnetic field, a high field magnetization process was measured at $T=1.4$ K. Obtained magnetization $M$ versus applied field $H$ and $dM/dH$ versus $H$ are shown in Fig. 3. The phase transitions are observed at $H_{\rm c1}=8.4$ T, $H_{\rm c2}=25.7$ T and $H_{\rm c3}=32.5$ T, which are indicated by arrows. We assign the field at which there is an inflection point in $dM/dH$ as the transition field. From the critical field $H_{\rm c1}$, the zero field gap $\Delta$ is evaluated as $\Delta/k_{\rm B}=g\mu_{\rm B}H_{\rm c1}/k_{\rm B}=11.2$ K. 
\begin{figure}
    \epsfigure{file=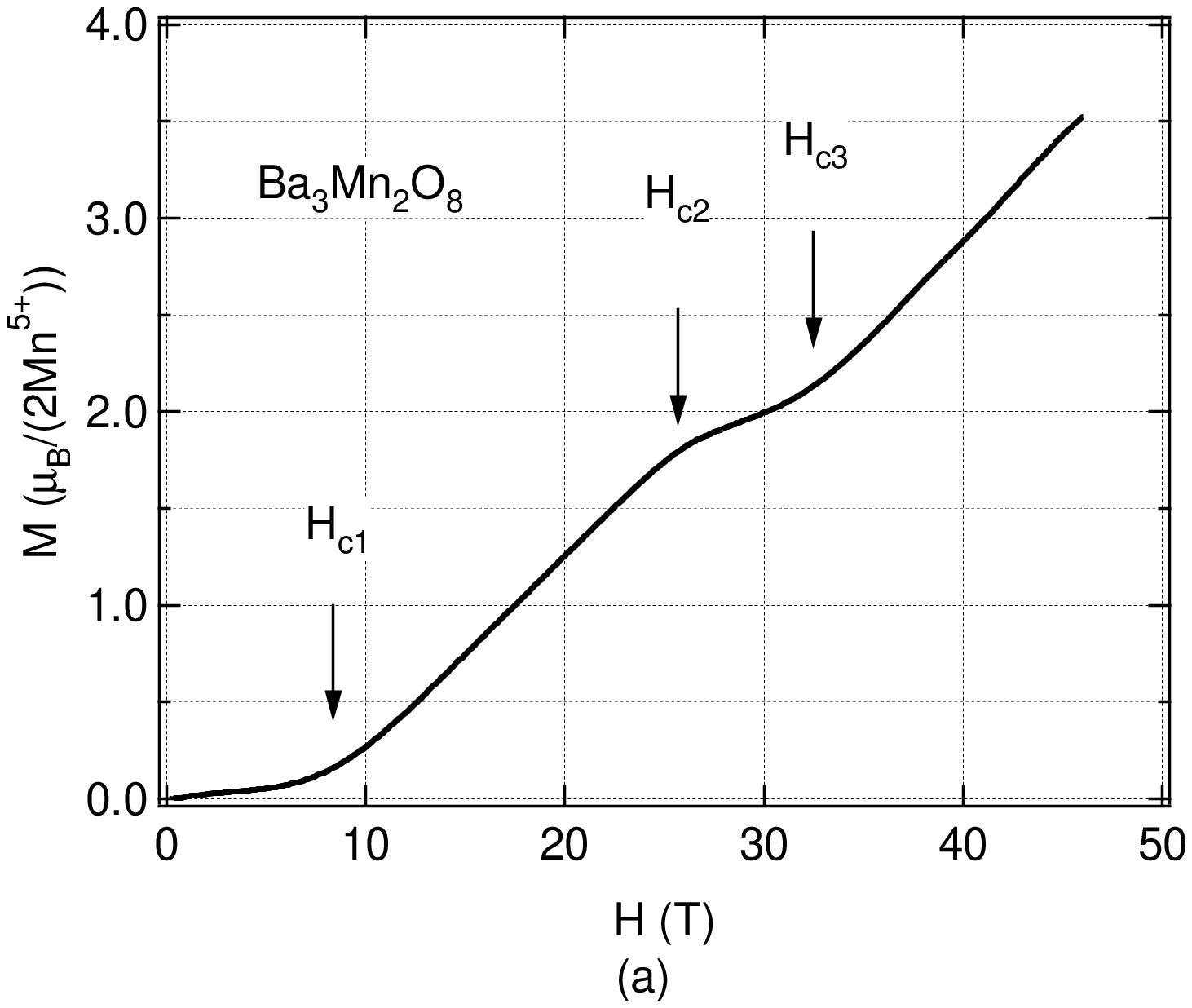,width=8.5cm}
    \epsfigure{file=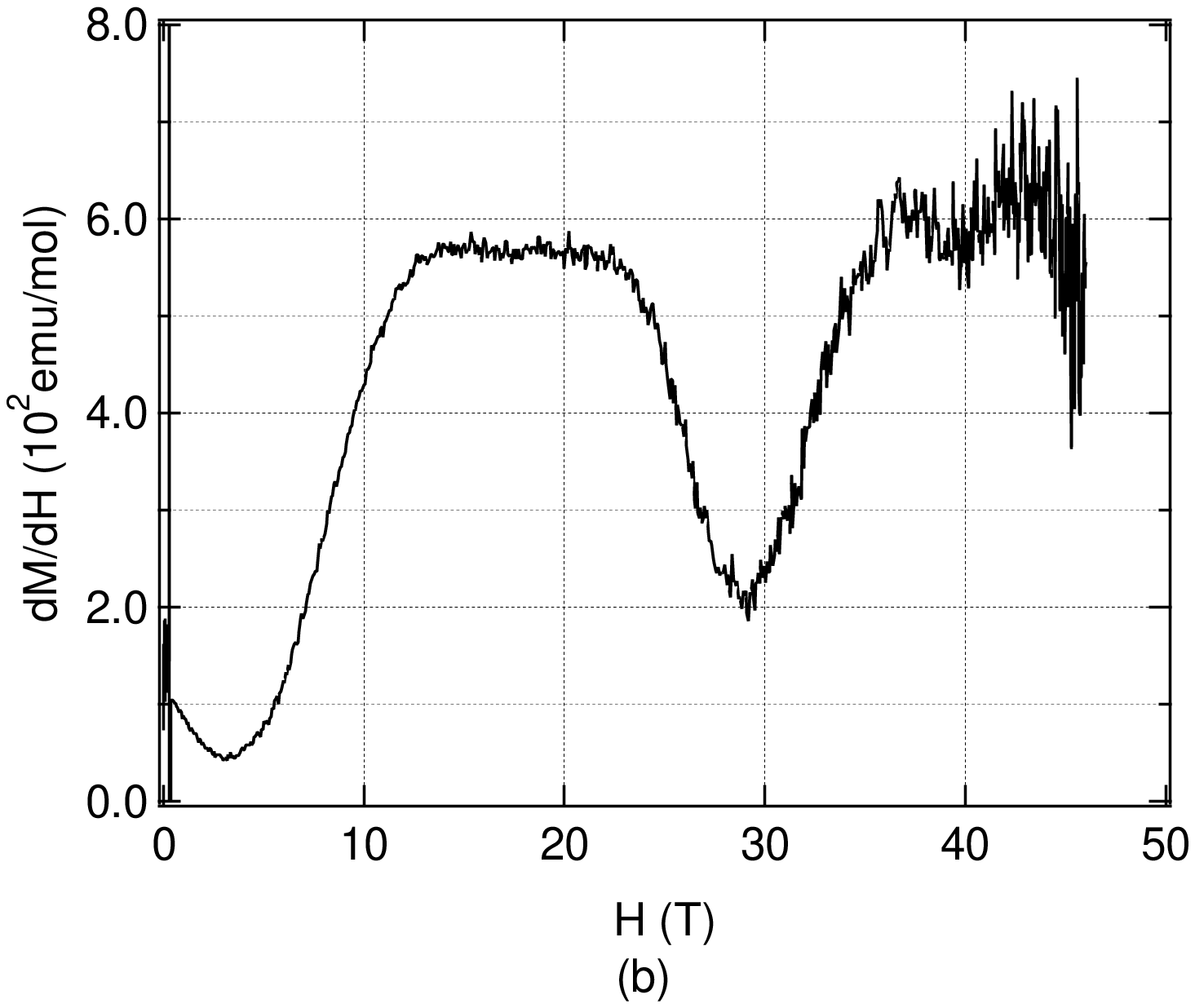,width=8.5cm}
    \caption{(a) Magnetization curve and (b) $dM/dH$ vs $H$ measured at $T=1.4$ K.}
    \label{high field}
\end{figure}

In the present measurements, the magnetization comes slightly short of saturation at the highest field of 47 T. Since the saturation magnetization per \mol{Mn^{5+}} ion can be estimated as $M_{\rm s}=g{\mu_{\rm B}}=1.98{\mu_{\rm B}}$, the saturation field is evaluated to be $H_{\rm s}\approx 50$ T. 

The notable feature of the magnetization curve is the two plateaus observed at $M=0$ and $M=\frac{1}{2}M_{\rm s}$. The edges of the plateaus are rounded due to the finite temperature effect, the magnetic anisotropy and the anisotropy of the $g$-factor.
The presence of these plateaus is consistent with the coupled $S=1$ dimer model, which was deduced from the susceptibility results. The plateaus at $M=0$ and $M=\frac{1}{2}M_{\rm s}$ correspond to $|0,0\rangle$ and $|1,1\rangle$, respectively, where $|S,S_z\rangle$ denotes the spin state of a dimer with the total spin $S$ and the $z$ component $S_z$. The slope regions observed for $H_{\rm c1}<H<H_{\rm c2}$ and $H_{\rm c3}<H<H_{\rm s}$ arise from the mixings between $|0,0\rangle$ and $|1,1\rangle$, and between $|1,1\rangle$ and $|2,2\rangle$, respectively, due to the interdimer interactions.\cite{Tachiki1,Tachiki2,Tsuneto} We note that the field ranges of two slope regions are almost the same, {\it i.e.,} $H_{\rm c2}-H_{\rm c1}\approx H_{\rm s}-H_{\rm c3}$.

In the field range, where the magnetization curve slopes, the ground state is gapless. Thus, magnetic ordering can occur with decreasing temperature, although no evidence of magnetic ordering in \green\ could be detected within the present measurements. The spin components perpendicular to the external field are relevant to the field-induced magnetic ordering. Recently, field-induced magnetic ordering has been attracting considerable attention.\cite{Chaboussant,Hammer,Honda,Manaka,Oosawa,Hagiwara,Nikuni} Nikuni {\it et al.}\cite{Nikuni} demonstrated that the field-induced magnetic ordering can be described by the Bose-Einstein condensation of magnons, when the system has the rotational symmetry around the magnetic field.  

From the exchange paths for $J_1$ and $J_2$ and the finding that $J_1+2J_2>0$, they both appear to be antiferromagnetic. The nature of the field-induced magnetic ordering in the present system depends strongly on the magnitudes of $J_1$ and $J_2$. 
The $J_1$ interaction stabilizes a structure in which the perpendicular spin components are arranged ferromagnetically on the triangular lattice and are stacked antiferromagnetically along the $c$-axis. In contrast, the $J_2$ interaction stabilizes a 120$^{\circ}$ structure on the triangular lattice for the perpendicular spin components. Consequently, the $J_1$ and $J_2$ interactions compete with each other. For this reason, the field-induced magnetic ordering in the present system is of great interest. 

The interdimer interactions will be determined from the dispersion relation of the magnetic excitations. Single crystals of \green\ are necessary to investigate the magnetic excitations. However, it is difficult to prepare \green\ in single crystal form, because this compound is decomposed at about $950^\circ$ C. An attempt to prepare single crystals is under way.

In conclusion, we have presented the results of magnetic susceptibility and high field magnetization measurements of \green\ polycrystal. We found that the magnetic ground state is a spin singlet with an excitation gap $\Delta/k_{\rm B}=11.2$ K. The magnetization curve has two plateaus at zero and at half of the saturation magnetization. These results can be understood by means of a coupled antiferromagnetic dimer model. The magnitudes of the intradimer and the interdimer interactions were evaluated to be $J_0/k_{\rm B}=9.2$ K and $(J_1+2J_2)/k_{\rm B}=6.6$ K, respectively.

\end{document}